\documentstyle[epsfig,12pt]{article}
\begin{document}
\newcommand{\gsi}{\,\raisebox{-0.13cm}{$\stackrel{\textstyle>}
{\textstyle\sim}$}\,}
\newcommand{\lsi}{\,\raisebox{-0.13cm}{$\stackrel{\textstyle<}
{\textstyle\sim}$}\,}
\newcommand{\be}{\begin{equation}} \newcommand{\ee}{\end {equation}}

\input epsf
\setlength{\oddsidemargin}{1.2 cm}
\setlength{\topmargin}{1.0 cm}
\setlength{\textwidth}{5.5 true in}
\setlength{\textheight}{8.0 true in}
\setlength{\parskip}{1.5 ex plus0.5ex minus 0.5ex}
\bibliographystyle{unsrt}
\def\question#1{{{\marginpar{\small \sc #1}}}}
\newcommand{\qqbar}{$q \bar{q}~$}
\newcommand{\slsh}{\rlap{$\;\!\!\not$}}     

\rightline{hep-ph/9701222}
\rightline{RAL-96-101}
\rightline{BHAM-HEP/96-05}  
\baselineskip=18pt
\vskip 0.7in
\begin{center}
{\bf \LARGE A Glueball-$q\bar{q}$ filter in Central Hadron Production}\\
\vspace*{0.9in}
{\large Frank E. Close}\footnote{\tt{e-mail: fec@v2.rl.ac.uk}} \\ 
\vspace{.1in}
{\it Rutherford Appleton Laboratory}\\
{\it Chilton, Didcot, OX11 0QX, England}\\ 
\vspace{0.1in} 
{\large Andrew Kirk}\footnote{\tt{e-mail: ak@hep.ph.bham.ac.uk}} \\ 
{\it School of Physics and Space Research}\\
{\it Birmingham University}\\
\end{center}

\begin{abstract}
We have stumbled upon a remarkable empirical feature of central meson production
which separates established $q\bar{q}$ mesons from glueball candidates.
This does not appear to have been noted previously and we have no simple
explanation for it. We suggest that glueballs and $q\bar{q}$ of the 
same $J^{PC}$ are distinguishable due to their boson versus fermion 
internal structure and that this leads to a different topology for 
central production of glueballs and $q\bar{q}$. Upon application
of this test to data from the WA102 experiment we find that
the $f_0(1500)$ and the $f_{2}(1900)$ show behaviour consistent with 
glueballs and opposite to that exhibited by established $q\bar{q}$ states.
\end{abstract}

\newpage
We propose a method for filtering glueballs with $J^{PC} = (0,1,2)^{++}$
from their $^3P_{(J=0,1,2)}$ $q \overline q$ 
counterparts when they are produced in the
central region of diffractive processes. The derivation is intuitive
rather than rigorous, yet its application to 
data from the CERN WA102 experiment turns out
to reveal some remarkable empirical regularities\cite{wa102new}.

These matters are timely given the
considerable current interest in the possibility that the lightest
glueball is a scalar with a mass of about $1.5$~GeV\cite{leap,landua96}.
 This is motivated both by
predictions of lattice QCD\cite{lattice}
 and by emerging hints in various experiments
where glueball production has historically been expected to be
 favoured\cite{landua96,expt,interfere,pdg96}.
One such mechanism is ``central production" where
the produced mesons have no memory of the
flavour of the initiating hadrons\cite{closerev}
 and are excited via the gluonic fields of
the ``Pomeron"\cite{landshoff}. Consequently it has been
anticipated that production of 
glueballs may be especially favoured in such processes.

However, such anticipation requires some caution.

First there is the well known problem that non-diffractive transfer
of flavour (Regge exchange) can
contaminate this simple picture and lead to the appearance of $q\bar{q}$
mesons in the central region. Furthermore, even for the diffractive production,
momentum transfer between the gluons of the Pomeron and 
the aligned constituents of the produced meson may
lead to either $gg$ or $q\bar{q}$ states. 
The former may be
favoured relative to $q\bar{q}$ production due to colour factors
 but unless further cuts are made to enhance the $gg$ signal, the appearance of
 novel states in central production is not of itself definitive evidence
for a glueball.

However,  there has been an interesting development with
the recent empirical observation\cite{wa9196} that the
states seen in central production are a function of the topology,
and depend on whether events are classified as either $LL$ or $LR$
(``left left" or ``left right" in the sense of how the beams scatter
into the final state relative to the initial direction). Specifically, when
the two beams scatter into opposing hemispheres ($LR$ as defined in
ref.\cite{wa9196}) the $f_1(1285)~ ^3P_1$ 
$q\bar{q}$ state is clearly visible (fig 1a) 
whereas in the same side configuration 
($LL$) it is less prominent relative to the structures in the
$1.4 - 2.0$~GeV mass range (fig 1b).
Such discrimination has also been seen for the $f_2(1270)$ 
$^3P_2$ $q\bar{q}$ state relative to the enigmatic $f_0(980)$ in the $\pi \pi$
channel\cite{wa9196}. Similar phenomena have recently been noted also
in Fermilab data \cite{fermi} and, in retrospect, at the ISR\cite{isr}.

This phenomenon led us to reconsider the  mechanisms for
the production of $gg$ and $q\bar{q}$ in the central region.
Our notation is that in the centre of mass frame
the initial protons have four-vectors
$p =(P+M^2/2P;p_T=0,p_L =P),
 ~ q =(P+M^2/2P;q_T = 0,q_L =-P)$,
the outgoing protons having
respectively momenta $p' \equiv (p'_L = x_a p; ~ p'_T)$ 
and $q' \equiv (q'_L = x_b q; ~ q'_T)$ 
and $x_F \equiv x_a - x_b$. 
The data have historically been
presented as a function of $M_R^2 \sim (1-x_a)(1-x_b)s$ with some
separation as a function of 
$t_a \sim -\frac{({p'_T})^2}{1-x_a}; 
~ t_b \sim -\frac{({q'_T})^2}{1-x_b}$. The topological separation into $LL$
and $LR$ is novel and independent of the magnitudes of $t_{a,b}$\cite{wa9196}.
We shall suggest that it is driven primarily by the variable
$dP_T \equiv |\vec{p'_T} - \vec{q'_T}|$ and that
$gg$ configurations are enhanced in kinematic
configurations where the gluons can flow ``directly" into the final state
with only small momentum transfer, in particular when $dP_T \to 0$.

A major uncertainty when analysing these processes is the modelling
of the Pomerons' interactions. While it
is now rather generally accepted that a Pomeron is a colour singlet
gluonic system (e.g. ref.\cite{landshoff}), 
an unknown feature is the topology of the 
individual constituent gluons in production processes.
For example,
Bialas and Landshoff\cite{landshoff} consider Higgs production by double
Pomeron exchange by
treating each Pomeron as a colour singlet system of two gluons; one gluon
is ``passive" and serves primarily to ensure overall 
colour singlet exchange (fig. 2a) while the other gluon in each Pomeron
transfers (longitudinal) momentum which stimulates the Higgs production.
This topology has some problems for the exclusive production of $q\bar{q}$
or a glueball. While it may be applicable to the pointlike Higgs,
for the 
exclusive production of a spatially extended hadron, which is the case of 
interest in the present paper,
one anticipates that it may be suppressed. The essential reason
is the large rescattering that is required to turn the large relative
longitudinal momenta of the ``active" gluons into co-moving constituents
necessary for the exclusive production of a composite hadron.

Specifically, consider the meson, $R$, to have an overall
longitudinal momentum $P \equiv P_3$ and to be made of 
constituents (e.g. gluons or quarks) of mass $m$.
The four momentum $p^{\mu}$ of a particle with mass $m$ may be
written
$p^{\mu} = (p^+ \equiv p_0 + p_3; p^- = \frac{m^2 + p_T^2}{p^+}; p_T)$.
The momentum of the meson $R$
is shared among its constituents, the $j$-th constituent
 carrying a fraction
$x_j = \frac{p_j^+}{P^+}$ where the total (time-like) momentum 
$P^+ = \Sigma_j p_j^+$ is conserved, $ \Sigma_j x_j =1$. The
relative momentum of the $j$-th constituent,
$l_j =  p_j - x_j P$,
is space like with $l^2 = - l_T^2$ and $l^+ = 0$. 
For a meson that is an $S$-wave state in its rest frame, the 
light cone wavefunction will have a structure that
may be parametrised in the form\cite{brodsky}
 $\phi_{LC}(x,l_T) \sim exp(-\Sigma
_j\frac{(l_T^2+m^2)R^2}{x_j})$ where the sum is over the constituents, $j$,
and $R$ is a measure of the hadron's size. 
(Excitation of
the $l_T$ degree of freedom corresponds to $P$-wave (and higher orbitals) in
the static limit\cite{weber}).
For a system of two equal mass constituents with $\delta x \equiv x_1-x_2$, 
the structure of the exponent becomes
$exp(-\frac{4(l_T^2+m^2)R^2}{1-\delta x^2})$ which is suppressed
if  $l_T$ is large or if $\delta x \to 1$.
This is essentially the well known form factor suppression of asymmetric
configurations\cite{brodsky,farrar}
 and becomes increasingly significant as $M_R$ becomes large.
It is manifested empirically in the sharp cut off of the $M_R$ distribution 
in central hadron production
(as for example when $M_R > 2$~GeV in fig 1
even though the kinematic reach of the experiment goes beyond this).

The above remarks are well known for the production of spatially extended
$q\bar{q}$ and will also be expected to apply to the production of glueballs.
However, in addition to these common features there can be generic
differences between the production of glueballs and $q \overline q$
of the same $J^{PC}$. For example,
a qualitative difference may result if glueballs are more pointlike
than $q\bar{q}$
or have a hard gluonic component in their wavefunction, either of
which would relatively enhance their production. The hard rescattering of
$gg \to gg$ rather than $gg \to q\bar{q}$ is also aided by colour factors.
There is also the possibility that the
more singular behaviour of propagators for confined gluons relative
to quarks at zero momentum
\cite{mrp} may lead to different production rates for
$q\bar{q}$ and $gg$ as a function of the kinematical variables,
for example in $dP_T$
and thereby in $LL$-$LR$.

The above description of the Pomeron 
is an extreme case in the sense that its gluons
are treated asymmetrically. The exclusive meson production then arises
when one ``hard" gluon from one Pomeron fuses with a gluon from the
other Pomeron. These two gluons have a large relative $p_L$
and in consequence are much separated in
$\delta x$; this disfavours exclusive production.
An alternative extreme, which can reduce this penalty, is where the two gluons
within a single Pomeron
cooperate in being strongly aligned, both in $p_L$ and $p_T$. 
In this case we can regard the initial proton
beam as a source of Pomerons, analogous to the way that an electron beam is 
a source of photons. Consider the process in the laboratory frame:
the Pomerons then scatter ``diffractively"
in the colour singlet gluonic field of the target (fig 2b) analogous to
the diffractive photoproduction of vector mesons. With the Pomeron
having $C=+$ we anticipate that this mechanism will
favourably produce $J^{PC} = 0^{++}, 2^{++}$
glueballs (which are the lightest according to lattice QCD). 
The Pomeron can in principle also convert into $J^{PC} = 0^{++},
2^{++}$ $q\bar{q}$ mesons in this process analogous to the 
photon turning into $J^{PC} = 1^{--}$ $q\bar{q}$. 
However, for $t$ and
$q'_T \to 0$, diffractive photoproduction
appears to produce only $^3S_1$ rather than $^3D_1$ configurations,
suggesting that the excitation of internal angular
momenta in the $q\bar{q}$
system is suppressed. If this is a guide to what happens with an
initial Pomeron ``beam" in place of a photon, then one may expect that
in this kinematic region
the $^3P_{0,2}$ $q\bar{q}$ will be disfavoured relative to their
glueball counterparts for which these $J^{PC}$ can be realised in $S$-wave.

The kinematics and experimental triggers cannot access this ``ideal"
situation and require $\vec{p'_T}$ and $\vec{q'_T}$ to be non-zero.
When the $\vec{p'_T}$ and
$\vec{q'_T}$ are co-moving and of equal magnitude such that $dP_T \to 0$, 
they tend to produce an overall transverse boost 
of the meson $R$ but with limited relative (internal) momentum: the
resulting configuration for 
$R$ will be strongly coupled to an $S$-state. 
By contrast, when the $\vec{p'_T}$ 
and $\vec{q'_T}$ are equal in magnitude but
anti-aligned (so $dP_T$ is large) then 
there can be significant relative $l_T$ ($\sim dP_T$) within $R$.
Note that in the laboratory frame $R$ is moving overall with large $p_L$
and that excitation of the $l_T$ degree of freedom corresponds to 
$P$-wave (and higher orbitals) in the static limit.
Thus by making the selection on data
that $dP_T \to 0$, there is the possibility that $q\bar{q}$ with 
$J^{PC} = 0^{++},1^{++},2^{++}$ (which are all $P$-wave composites) will
be suppressed relative to glueballs (or at least, 
relative to $S$-wave bound states of bosons with these $J^{PC}$).
In effect, the Bialas~-~Landshoff topology encounters the penalty of mismatched 
$p_L$, which disfavours exclusive meson production at large $M_R$; 
the alternate topology
dilutes this problem in $p_L$ but in $p_T$ it remains. It is this which
is varied $via$ $dP_T$ and which can govern the relative importance of
$S$ and $P$ (or higher) wave states.

If this is realised then we expect $f_1(1285)$, $f_2(1270)$ $inter$ $ alia$ to
be suppressed as $dP_T \to 0$;
we expect $f_0~ ^3P_0$ $q\bar{q}$ to be suppressed
similarly whereas $f_0(1500) \equiv gg$ or $f_0(980) \equiv K\bar{K}$ would
be able to survive. Conversely, at larger $dP_T$ the production of
$gg$ and $q\bar{q}$ can be competitive, their relative production
rates being dependent on
their detailed internal wavefunctions and the relative importance
of the colinear versus asymmetric (hard gluon) production mechanisms.

 The above intuitive picture is only a first sketch of the
full dynamics at best but at least it provides a starting point that is
qualitatively consistent with the
pattern of the $LL$ and $LR$ topologies.
Intuitively one expects that the $LL$ configuration will contain a large 
sample with comparatively small relative transverse momentum and hence $dP_T
 \to 0$ favouring the glueballs.
In contrast,
the $LR$ topology will have a tendency for rather larger 
relative transverse momentum, thereby necessitating greater momentum transfer
in the exclusive meson production vertex; thus for this
case one will expect both $gg$ or $q\bar{q}$ production to occur. The
relative emergence of the glueball candidates compared to the established
$f_1(1285)$ $q\bar{q}$ in fig.~1 are in accord with this. However, if 
it is the $dP_T$ variable that underpins the $LL-LR$ effect,
then there should be a more marked effect when
the data are selected directly as a function of
$dP_T = |\vec{p'_T} - \vec{q'_T}|$. 
Accordingly we have done so and present the results for the $4 \pi$ channel
in fig(3).

The  $dP_T \geq 0.5$ GeV (fig 3a) is similar to the original $LR$
sample as expected. The sample with 0.2~GeV~$\leq$~$dP_T$~$\leq$~0.5~GeV 
(fig 3b) shows the $f_1(1285)$ becoming suppressed and a sharpening of the
$f_0(1500)$ and $f_{2}(1900)$ structures. However, the most
dramatic effect is seen in the  $dP_T$~$\leq$~0.2~GeV sample (fig 3c) where the
$f_1(1285)$, a $q\bar{q}$ state, has essentially disappeared while the 
$f_0(1500)$ and $f_{2}(1900)$ structures have become more clear. 
(This is driven by $dP_T \to 0$ and is not an artifact of
Bose symmetry, as might have occurred if the Pomeron - photon analogy were 
exact: if we select $t_1 \neq t_2$, the $f_1(1285)$ still disappears as
$dP_T \to 0$).
These surviving structures have been identified as 
glueball candidates: the $f_0(1500)$ is motivated by lattice QCD while
the $f_2(1900)$ is noted to have the right mass to lie on the Pomeron
trajectory\cite{pvl}.

The $f_0(1500)$ is rather clean and
appears at $dP_T \to 0$ with a 
shape and mass that are not inconsistent with what is seen in
$p\bar{p}$ annihilation.
This is in contrast to the full data sample of the present
experiment where this state interfered with the $f_0(1370)$ and was
shifted to a lower mass ($\sim 1440$ MeV)\cite{interfere} and
with a much narrower width ($\sim 60$ MeV). The emergence in fig 2c of a more
canonical $f_0(1500)$ \cite{pdg96}
suggests, at least implicitly, that the
$f_0(1370)$ $q\bar{q}$\cite{cafe95} state has become suppressed as
$dP_T \to 0$ while the ($gg$ candidate) $f_0(1500)$ has survived. It is
important that experiments now verify if this is indeed the case.

Similar cuts have been applied to the $\pi\pi$, $K\bar{K}$ and $K\bar{K} \pi$
data (see figs 4 et seq. taken from ref.\cite{wa102new}).
 The $f_1(1285)$ and $f_1(1420)$ are seen in the channel $K\bar{K} \pi$
when $dP_T > 0.5 ~ GeV$
but vanish when $dP_T \to 0$ in line with $^3P_1$ $q\bar{q}$. We note
that there are no $1^{++}$
resonances visible in this limit, in line with
lattice QCD which predicts that there are no $1^{++}$ glueballs below 
$\sim 3.5$~GeV~\cite{teper}. 
In the $\pi \pi$ channel we see the $f_2(1270)$ when 
$dP_T > 0.5$~GeV. This vanishes as $dP_T \to 0$ as expected for
$^3P_2$ $q\bar{q}$. 
 The survival of $f_0(980)$ is significant and we suggest 
that this shows its affinity for coupling via $S-$wave bosons. This could
be due to a $gg$ presence or to $K\bar{K}$ in its wavefunction\cite{wein,torn};
at the present level of analysis we are unable to distinguish between these
alternatives.

To summarise: we have stumbled upon a remarkable feature of central meson
production that does not appear to have been noticed previously. Although
its extraction via the $dP_T$ cut was inspired by intuitive arguments
following the observation of an $LL-LR$ asymmetry, we have no
simple dynamical explanation. Nonetheless, the empirical message is
dramatic enough to stand alone and thereby
we suggest that a systematic study of meson production as
a function of $dP_T \equiv |\vec{p'_T} - \vec{q'_T}|$ holds special
promise for isolating the systematics of meson production in the central region
and in filtering $q\bar{q}$ mesons from those with significant $boson$-$boson$
content. The latter include $K\bar{K}$ molecular bound states (or $s\bar{s}$
states with significant $K\bar{K}$ component in the wavefunction), $pomeron$
- $pomeron$ states and $glueballs$.
Our selection
procedure will need to be tested  further in future experiments in
order to determine  the extent of its empirical validity. In turn we
hope  thereby that its dynamical
foundations may be put on a sounder footing and the filtering of glueballs be
made a practical reality.

\vskip 0.4cm
\hspace*{2em}
FEC is partially
supported by the European Community Human Mobility Program Eurodafne,
Contract CHRX-CT92-0026
and is grateful to the WA102 collaboration for this interaction.

\newpage

\begin{figure}
\epsfxsize=\hsize
\epsffile{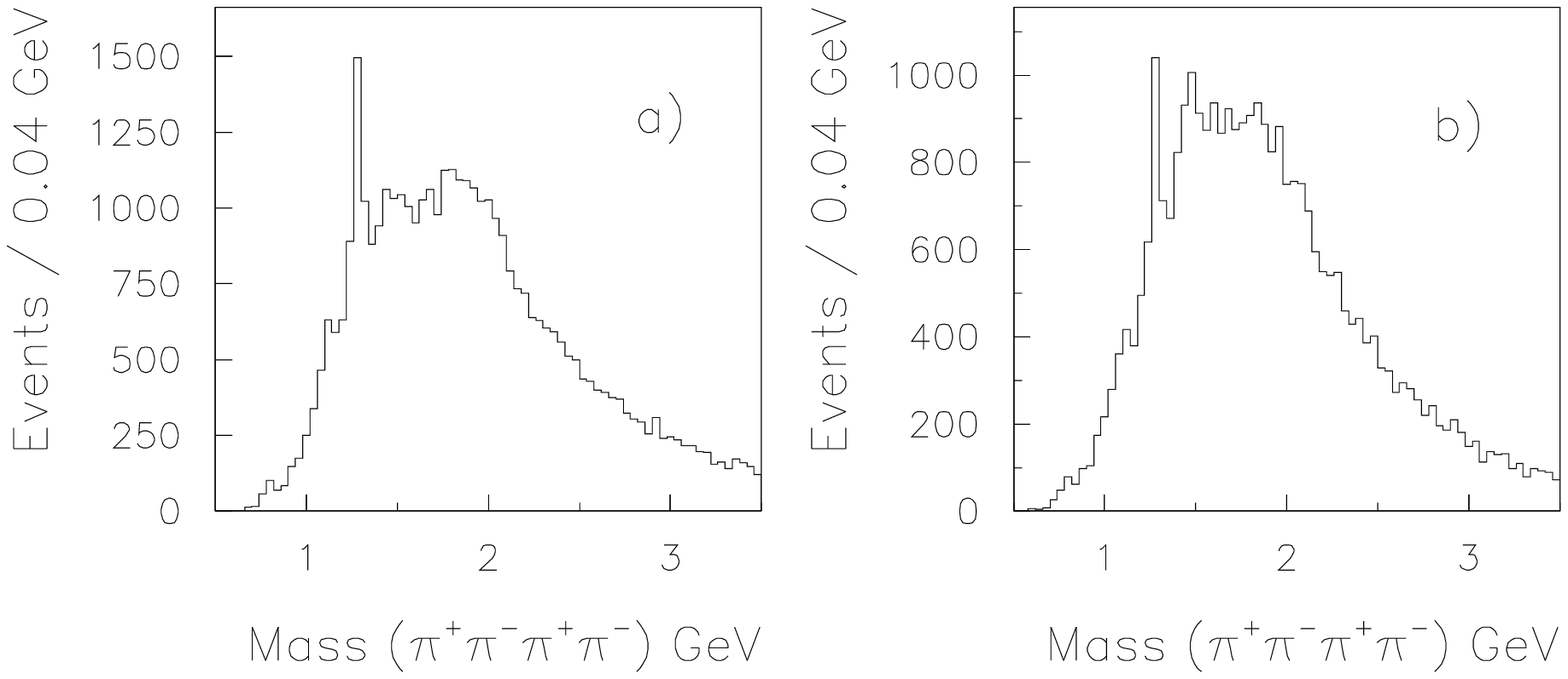}
\caption{The 4$\pi$ mass spectra from ref.10 for (a) LR and (b) LL
topologies.}  
\label{lllr}
\end{figure}
\begin{figure}
\epsfxsize=\hsize
\epsffile{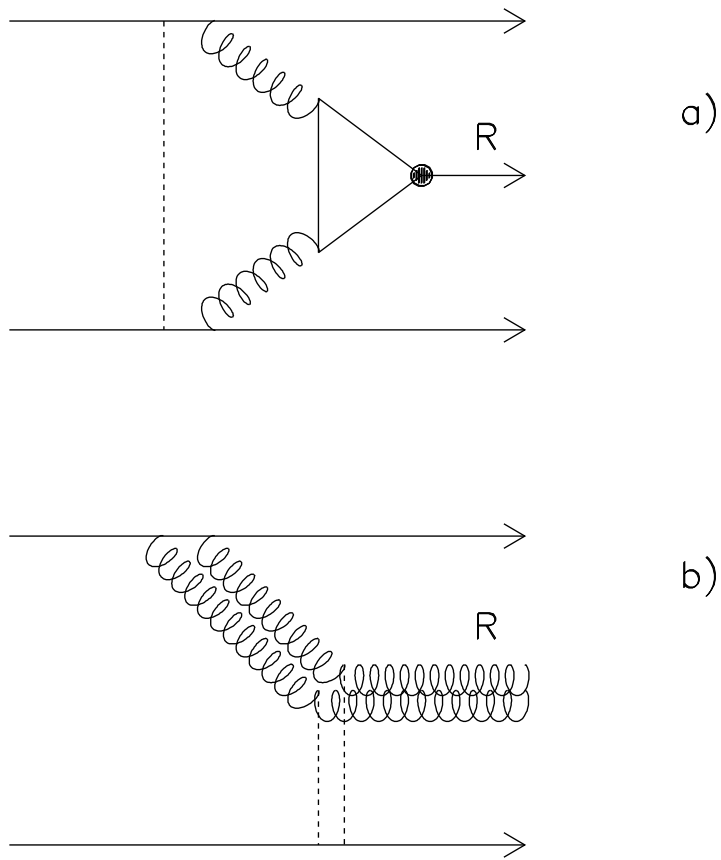}
\caption{ (a) Two gluons with large $p_L$ fuse to make a meson $R$. 
(b) Diffractive scattering of a gluonic Pomeron to produce a glueball}
\label{fi:1}
\end{figure}
\begin{figure}
\epsfxsize=\hsize
\epsffile{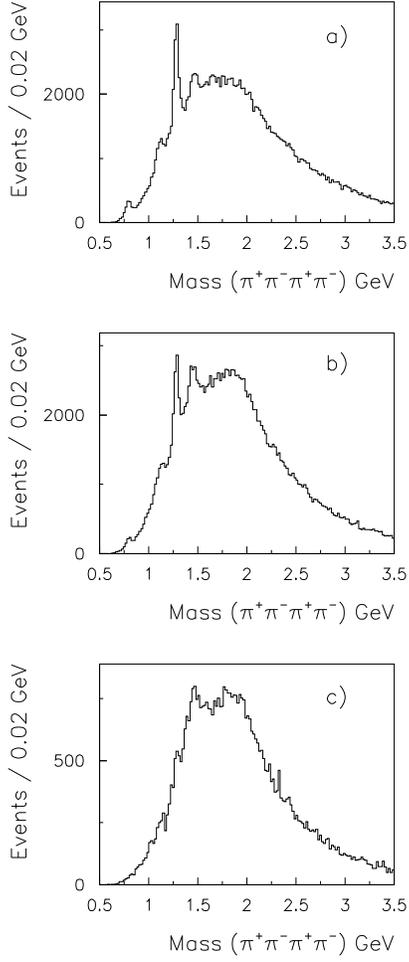}
\caption{The 4$\pi$ mass spectra (i) With $dP_T > 0.5$~GeV
exhibiting a clear $f_1(1285)$; (ii) $0.2 < dP_T < 0.5$~GeV
(iii) $dP_T < 0.2$~GeV where the $f_1(1285)$ has disappeared while
the $f_0(1500)$ is seen more clearly.}
\label{4pi}
\end{figure}
\begin{figure}
\epsfxsize=\hsize
\epsffile{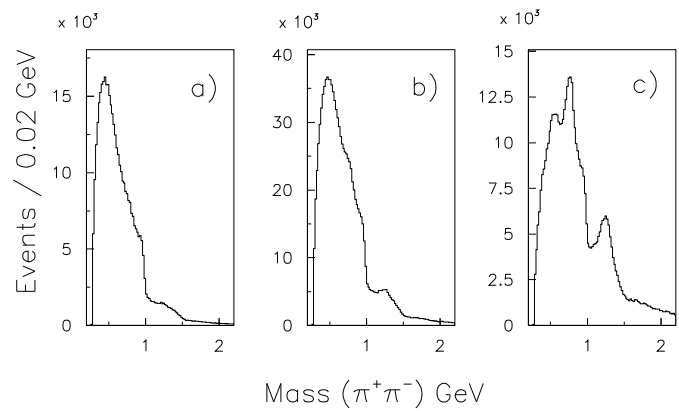}
\caption{The $\pi\pi$ mass spectrum for 
c) $dP_T<0.2$ GeV, d) $0.2 <dP_T <   0.5$ GeV and e) $dP_T >   0.5$ GeV.}
\label{fi:2pi}
\end{figure}
\begin{figure}
\epsfxsize=\hsize
\epsffile{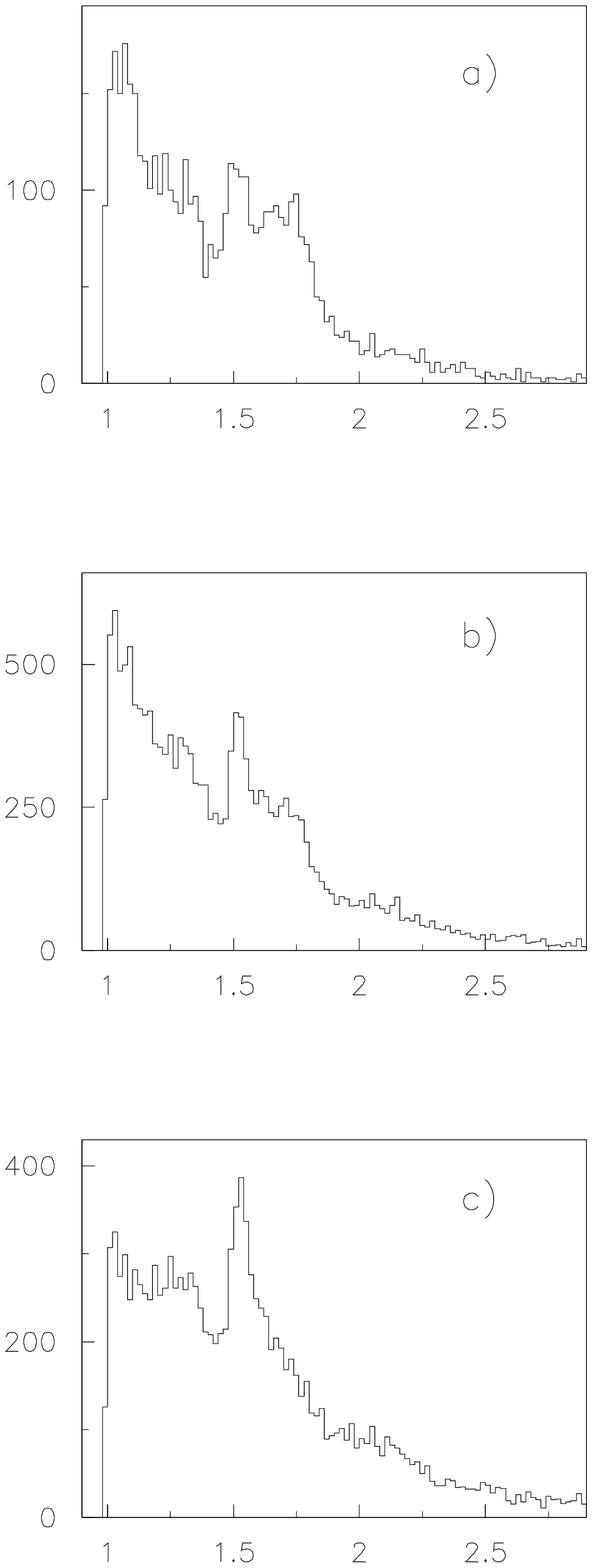}
\caption{$K^+K^-$ mass spectrum for a) $dP_T <   0.2$ GeV, 
b) $0.2 <   dP_T <   0.5$ GeV and c) $dP_T >   0.5$ GeV}
\label{fi:2k}
\end{figure}

\end{document}